\documentstyle[mprocl]{article}

\input{psfig}
\def\physrep{{\em Phys. Rep.}}
\def\apj{{\em Ap. J.}}
\def\apjl{{\em Ap. J. Lett.}}
\def\npa{{\em Nucl. Phys.} A}

\def\plb{{\em Phys. Lett.}  B}
\def\prl{{\em Phys. Rev. Lett.}}
\def\prc{{\em Phys. Rev.} C}
\def\prd{{\em Phys. Rev.} D}

\def\be{\begin{equation}}
\def\ee{\end{equation}}
\def\bea{\begin{eqnarray}}
\def\eea{\end{eqnarray}}
\newcommand \ga{\raisebox{-.5ex}{$\stackrel{>}{\sim}$}}
\newcommand \la{\raisebox{-.5ex}{$\stackrel{<}{\sim}$}}

\begin{document}

\title{ NEUTRON STARS: RECENT DEVELOPMENTS}

\author{HENNING HEISELBERG}

\address{Nordita, Blegdamsvej 17,\\2100 Copenhagen \O, Denmark\\
E-mail: hh@nordita.dk}

\maketitle
\abstracts{Recent developments in neutron star theory and observation
are discussed.  Based on modern
nucleon-nucleon potentials more reliable equations of state for dense
nuclear matter have been constructed.  Furthermore, phase transitions
such as pion, kaon and hyperon condensation, superfluidity and quark
matter can occur in cores of neutron stars. Specifically, the nuclear
to quark matter phase transition and its mixed phases with intriguing
structures is treated.  Rotating neutron stars with and without phase
transitions are discussed and compared to observed masses, radii and
glitches.  The observations of possible heavy $\sim 2M_\odot$ neutron
stars in X-ray binaries and QPO's require relatively stiff equation of
states and restrict strong phase transitions to occur at very high
nuclear densities only.}

\section{Introduction}

Neutron stars are complicated many-body systems of neutrons, protons,
electrons and muons with possibly additional intricate phases of pion, kaon,
hyperon or quark condensates.  It is therefore appropriate
that neutron stars are included at this ``Xth Intl. Conf. on  Recent Progress
in Many-body Theories''.

A brief history of the most important discoveries in this millennium
concerning neutron stars listed
in Table I. Most are well known except perhaps for
the most recent ones discussed the following sections:
Sec. 2 list neutron star masses from binary pulsars and X-ray binaries
and specifically discuss recent masses and radii from quasi periodic
oscillations in low mass X-ray pulsars.
In Sec. 3 we turn to modern equation of states for neutron star
matter with particular attention to 
the uncertainty in the stiffness of the equation of
state at high densities and causality constraints. Sec. 4 
attempts to give an up-to-date on possible phase transitions
to kaon and pion condensates, hyperon and quark matter, superfluidity, etc.
Sec. 5 contains the resulting structure of neutron stars,
calculated masses and radii, and compares to observations.
In Sec. 6 observational effects of glitches are described
when phase transitions occur.
Sec. 7 contain novel information on connections between supernova
remnants and $NO^-_3$ peaks in ice cores, X-ray bursts and thermonuclear
explosions, gamma ray bursters and neutron star collapse to black holes.
Finally, a summary and conclusion is given. For more details we refer to
\cite{physrep}.

\begin{table}[htp]
\vspace{-1cm}
\begin{center}
\caption{Chronological list of important developments related to 
neutron stars}         
\begin{tabular}{rll} \\\hline\noalign{\smallskip}
     Year & ``Observers''    & Discovery \\
\noalign{\smallskip}\hline\noalign{\smallskip}
     1054 & Chinese          & record the Crab Supernova   \\
     1572 & Tycho Brahe      & observes a Supernova    \\
     1932 & Chadwick         & discovers the neutron  \\
     1932 & Landau \& others & suggests the existence of neutron stars\\
     1934 & Baade \& Zwicky  & connects supernovae to gravitational\\
          &                  &  collapse of stars to neutron stars\\
     1935 & Oppenheimer \& Volkoff  & calculate first
            neutron star structures\\
     1946 & Gamow            & develops nucleosynthesis which \\
          &                  & requires heavier elements from Supernovae\\
     1967 & Bell \& Hewish   & discover first pulsar  \\
     1969 &          & pulsars in Crab and Vela Supernova remnants \\
     1973 & Hulse \& Taylor  & discover first binary pulsar  \\
     1987 & Neutrino detectors & collect 19 neutrinos from SN-1987A  \\
     1995 & Nijmegen data base & compilation of $\ga5000$ NN cross sections\\
          &                  & leads to ``modern'' NN potentials and EoS\\
     1996 & RXTE  & kHz oscillations (QPO) in X-ray binaries \\
     1997 & BeppoSAX & Gamma Ray Burst with afterglow at $z\ga1$ \\
          &          & Beaming and syncroton radiation observed \\
     1998 & Supernova remnants & match dating of polar ice core 
            $NO^-_3$-peaks\\
\noalign{\smallskip}\hline
\label{history}
\end{tabular}
\end{center}
\vspace{-0.3cm}
\end{table}

\section{Observed neutron star masses and QPO's}

The best determined neutron star masses are found in binary pulsars
and all lie in the range $1.35\pm 0.04 M_\odot$ \cite{Thorsett}.
These masses have been accurately determined from
variations in their radio pulses due to doppler shifts as well
periastron advances of their close elliptic orbits that are
strongly affected by general relativistic effects.
One exception is the nonrelativistic pulsar PSR J1012+5307
of mass\footnote{All uncertainties given here
are 95\% conf. limits or $\sim2\sigma$}
$M=(2.1\pm 0.8)M_\odot$ \cite{Paradijs}. 

Several X-ray binary masses have been measured of which the heaviest
are Vela X-1 with\cite{Barziv} $M=(1.9\pm 0.2)M_\odot$ and Cygnus X-2
with\cite{Orosz} $M=(1.8\pm 0.4)M_\odot$.  Their Kepler orbits are
determined by measuring doppler shift of both the X-ray binary and its
companion. To complete the mass determination one needs the orbital
inclination which is determined by eclipse durations, optical light
curves, or polarization variations\cite{Paradijs}.

The recent discovery of high-frequency brightness oscillations in
low-mass X-ray binaries provides a promising new method for
determining masses and radii of neutron stars\cite{Lamb}. The
kilohertz quasi-periodic oscillations (QPO) occur in pairs and are
most likely the orbital frequencies
\bea
   \nu_{QPO}=(1/2\pi)\sqrt{GM/R_{orb}^3} \,, \label{orb}
\eea
of accreting matter
in Keplerian orbits around neutron stars of mass $M$ and its beat
frequency with the neutron star spin, $\nu_{QPO}-\nu_s$. 
The accretion
can for a few QPO's be tracked to its innermost stable orbit 
\cite{Zhang,Kaaret}
\bea
   R_{ms} &=& 6GM/c^2 \,. \label{iso}
\eea
 For slowly rotating stars the resulting mass is from 
Eqs. (\ref{orb},\ref{iso})
\bea
    M &\simeq& 2.2M_\odot \frac{{\mathrm{kHz}}}{\nu_{QPO}}  \,.
\eea
  For example, the
maximum frequency of 1060 Hz upper QPO observed in 4U 1820-30 gives $M\simeq
2.25M_\odot$ after correcting for the $\nu_s\simeq275$ Hz neutron star
rotation frequency.  If the maximum QPO frequencies of 4U 1608-52
($\nu_{QPO}=1125$ Hz) and 4U 1636-536 ($\nu_{QPO}=1228$ Hz) also
correspond to innermost stable orbits the corresponding masses are
$2.1M_\odot$ and $1.9M_\odot$.  
Evidence for the innermost stable orbit has been found
for 4U 1820-30, where $\nu_{QPO}$ display a distinct saturation
with accretion rate indicating that orbital frequency cannot 
exceed that of the innermost stable orbit.
However, one caveat is that large accretion leads to radiation that
can slow down the rotation of the accreting matter.
More observations are needed before a firm conclusion can
be made.

Large neutron star masses of order $\sim 2M_\odot$ would  restrict
the equation of state (EoS) severely for dense matter as addressed in the
following.

\section{Modern nuclear equation of states}

Recent models for the nucleon-nucleon (NN) interaction, based on the
compilation of more than 5000 NN cross sections in the Nijmegen data
bank, have reduced the uncertainty in NN potentials.  
The last Indiana run at higher momenta will further reduce uncertainties
in NN interactions.
Including
many-body effects, three-body forces, relativistic effects, etc., the
nuclear EoS have been constructed with reduced uncertainty allowing
for more reliable calculations of neutron star properties \cite{Akmal}.
Likewise, recent realistic effective interactions for nuclear matter
obeying causality at high densities, constrain the EoS
severely and thus also the maximum masses of neutron stars. We
have in \cite{physrep} elaborated on these analyses by incorporating causality
smoothly in the EoS for nuclear matter allowing for first and second
order phase transitions to, e.g., quark matter.

For the discussion of the gross properties of neutron stars we will
use the optimal EoS of Akmal, Pandharipande, \& Ravenhall \cite{Akmal}
(specifically the Argonne $V18 + \delta v +$ UIX$^*$ model- hereafter
APR98), which is based on the most recent models
for the nucleon-nucleon interaction, see Engvik et al. \cite{Engvik} for a
discussion of these models, and with the inclusion of a parametrized
three-body force and relativistic boost corrections. The EoS for
nuclear matter is thus known to some accuracy for densities up to a
few times nuclear saturation density $n_0=0.16$ fm$^{-3}$.  We
parametrize the APR98 EoS by a simple form for the compressional and
symmetry energies that gives a good fit
around nuclear saturation densities and smoothly incorporates
causality at high densities such that the sound speed approaches the
speed of light.
This requires that the compressional part of the
energy per nucleon is quadratic
in nuclear density with a minimum at saturation but linear at high densities
\begin{eqnarray}
    {\cal E} &=& E_{comp}(n) + S(n)(1-2x)^2 \nonumber\\
   &=& {\cal E}_0 u\frac{u-2-s}{1+s u} +S_0 u^\gamma (1-2x)^2.   
    \label{eq:EA} 
\end{eqnarray}
Here, $n=n_p+n_n$ is the total baryon density, $x=n_p/n$ the
proton fraction and $u=n/n_0$ is the ratio of the baryon density to
nuclear saturation density. The compressional term is in Eq.\
(\ref{eq:EA}) parametrized by a simple form which reproduces the
saturation density and the binding energy per
nucleon ${\cal E}_0=15.8$MeV at $n_0$ of APR98. The ``softness''
parameter $s\simeq 0.2$, which gave the best fit to the data
of APR98   
is determined \cite{physrep} by fitting the energy per
nucleon of APR98 up to densities of $n\sim 4n_0$.
For the symmetry energy
term we obtain $S_0=32$ MeV and $\gamma=0.6$ for the best fit. The
proton fraction is given by $\beta$-equilibrium at a given density.

The one unknown parameter $s$ 
expresses the uncertainty in the EoS at high
density and we shall vary this parameter within the allowed limits in
the following with and without phase transitions to calculate mass,
radius and density relations for neutron stars.
The ``softness'' 
parameter $s$ is related to the incompressibility
of nuclear matter as $K_0=18{\cal E}_0/(1+s)\simeq 200$MeV. It 
agrees with the poorly known experimental value \cite{Blaizot}, 
$K_0\simeq 180-250$MeV which does not restrict it
very well.  From $(v_s/c)^2=\partial P/\partial (n\cal{E})$, where $P$ is the
pressure, and the EoS  of Eq.\ (\ref{eq:EA}),
the causality condition $c_s\le c$ requires 
\begin{equation}
      s \ga \sqrt{\frac{{\cal E}_0}{m_n}} \simeq 0.13 \,,\label{causal}
\end{equation}
where $m_n$ is the mass of the nucleon.
With this condition we have a causal EoS that reproduces the
data of APR98 at densities up to $0.6\sim 0.7$ fm$^{-3}$. 
In contrast, the EoS of APR98 becomes
superluminal at $n\approx 1.1$ fm$^{-3}$.  For larger $s$ values
the EoS is softer which eventually leads to smaller maximum masses of
neutron stars. The observed $M\simeq 1.4M_\odot$ in binary pulsars
restricts $s$ to be less than $0.4-0.5$ depending on rotation
as shown in calculations of neutron stars
below. 

\begin{figure}
\vskip -6cm
{\centering\mbox{\psfig{file=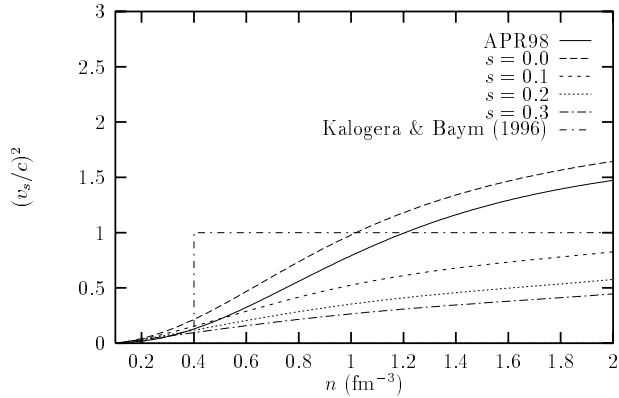,height=20cm,angle=0}}}
\vskip -7cm
\caption{$(v_s/c)^2$ for $\beta$-stable $pn$-matter for $s=0,0.1,0.2,0.3$, 
the results of APR98, and for the patched EoS of Kalogera \& Baym 
which shows a discontinuous $(v_s/c)^2$.\label{fig1}  } 
\end{figure}

In Fig.\ \ref{fig1} we plot the sound speed $(v_s/c)^2$ for
various values of $s$ and that resulting from the microscopic
calculation of APR98 for $\beta$-stable $pn$-matter.  The form of
Eq.~(\ref{eq:EA}), with the inclusion of the parameter $s$, provides
a smooth extrapolation from small to large densities
such that the sound speed $v_s$ approaches the
speed of light. For $s=0.0$ ($s=0.1$) the EoS
becomes superluminal at densities of the order of 1 (6) fm$^{-3}$.

The sound speed of Kalogera \& Baym \cite{Kalogera} is also plotted in Fig.\
\ref{fig1}. It jumps discontinuously to the speed of light at a
chosen density. With this prescription they were able to obtain an
optimum upper bound for neutron star masses and obey causality.  This
prescription was also employed by APR98.
The EoS is thus discontinuously stiffened by
taking $v_s=c$ at densities above a certain value $n_c$ which,
however, is lower than $n_{s}=5n_0$ where their nuclear EoS becomes
superluminal. This approach stiffens the nuclear EoS for densities
$n_c<n<n_s$ but softens it at higher densities. Their resulting
maximum masses lie in the range $2.2M_\odot\la M\la 2.9M_\odot$.  Our
approach however, incorporates causality by reducing the sound speed
smoothly towards the speed of light at high densities. Therefore our
approach will not yield an absolute upper bound on the maximum mass
of a neutron star
but gives reasonable estimates based on modern EoS around nuclear matter
densities, causality constraints at high densities and a smooth
extrapolation between these two limits (see Fig. \ref{fig1}).

At very high densities particles are expected to be relativistic 
and the sound speed should be smaller than the speed of light,
$v_s^2\simeq c^2/3$. Consequently, the EoS should be even softer at
high densities and the maximum masses we obtain with the EoS
of (\ref{eq:EA}) are likely to be too high estimates.

\section{Phase transitions}

 The physical state of matter in the interiors of neutron stars at
densities above a few times normal nuclear matter densities is
essentially unknown and many first and second order phase transitions
have been speculated upon.

\subsection{Kaon condensation}

 Kaon condensation in dense matter was suggested by Kaplan and Nelson
\cite{kn87}, and has been discussed in many recent publications
\cite{BLRT,Weise}. Due to the attraction between $K^-$ and
nucleons its energy decreases with increasing density, and eventually
if it drops below the electron chemical potential in neutron star matter in
$\beta$-equilibrium, a Bose condensate of $K^-$ will appear.
It is found that $K^-$'s condense at densities above
$\sim 3-4\rho_0$, where $\rho_0=0.16$ fm$^{-3}$ is normal nuclear matter
density. This is to be compared to the central density of
$\sim4\rho_0$ for a neutron star of mass 1.4$M_\odot$ according to the
estimates of Wiringa, Fiks and Fabrocini \cite{WFF} using realistic
models of nuclear forces. 

In neutron matter at low densities, when the interparticle spacing is much
larger than the range of the interaction, $r_0\gg R$,
the kaon interacts strongly many times with the same nucleon before it
encounters and interacts with another nucleon. 
Thus one can use the scattering length as the ``effective'' kaon-neutron
interaction, $a_{K^-N}\simeq -0.41$fm, where we ignore the minor proton
fraction in nuclear matter. The kaon energy deviates from its 
rest mass by the Lenz potential
\begin{equation}
   \omega_{Lenz} = m_K + \frac{2\pi}{m_R}\, a_{K^-N}\, n_{NM} ,
               \label{Lenz}
\end{equation}
which is the optical potential obtained in the impulse 
approximation. If hadron masses furthermore 
decrease with density the condensation will occur at lower densities
\cite{BLRT}.

\begin{figure}
\begin{center}
\vspace{-1cm}
{\centering
\mbox{\psfig{file=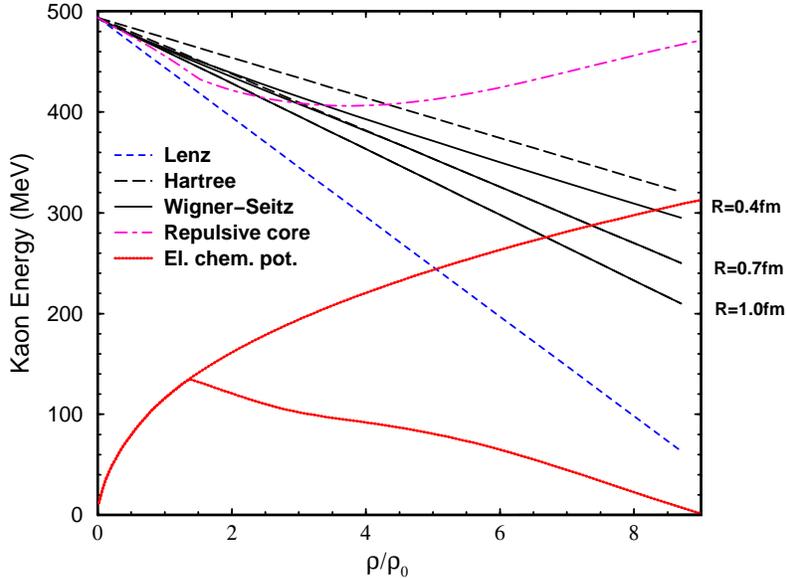,height=9cm,angle=-90}}}
\caption{ Kaon energy as function of neutron density is shown by full
curves for various ranges of the $K^-n$ potentials $R=0.4, 0.7, 1.0$ fm. 
Nuclear correlations change the low density Lenz
result (Eq. (\ref{Lenz}), dotted curve) to the high density Hartree
result (Eq.(\ref{Hartree}), dashed curves).  The electron chemical
potential $\mu_e$ from the EoS of Eq.
(\ref{eq:EA}) with $\delta=0.2$ is
shown with and without (lower and upper dotted curves) a transition to
a mixed phase of quark matter for $B=100$
MeVfm$^{-3}$. \label{fig:kaoncondens} }
\end{center}
\vspace{-0.5cm}
\end{figure}

At high densities when the interparticle spacing is much
less than the range of the interaction, $r_0\ll R$, the kaon
will interact with many nucleons on a distance scale much less than
the range of the interaction. 
The kaon thus experiences the field from many nucleons
and the kaon energy deviates from its rest mass 
by the Hartree potential:
\begin{equation}
   \omega_{Hartree} = m_K + n_{NM} \int V_{K^-N}(r)
           d^3r          \,,\label{Hartree}
\end{equation}
As shown in Ref.\ \cite{PPT}, the Hartree potential is considerably less
attractive than the Lenz potential.  Already at rather low densities,
when the interparticle distance is comparable to the range of the $KN$
interaction, the kaon-nucleon and nucleon-nucleon correlations
conspire to reduce the $K^-N$ attraction significantly \cite{PPT,Carlson}.
This is also evident from Fig.\ \ref{fig:kaoncondens} 
where the transition from the low
density Lenz potential to the high density Hartree potential is
calculated by solving the Klein-Gordon equation for kaons in neutron
matter in the Wigner-Seitz cell approximation. Results are for square
well $K^-N$-potentials of various ranges $R$.
For the measured $K^-n$
scattering lengths and reasonable ranges of interactions the
attraction is reduced by about a factor of 2-3 in cores of neutron
stars. Relativistic effects further reduce the attraction at high
densities. Consequently, a kaon condensate is less likely in neutron
stars due to nuclear correlations.

If the kaon condensate occurs a mixed phase of kaon condensates and
ordinary nuclear matter may coexist in a mixed phase \cite{Schaffner}
depending on surface and Coulomb energies involved. The structures
would be much like the quark and nuclear matter mixed phases described
above.

\subsection{Pion condensation}

Pion condensation is like kaon condensation possible
in dense neutron star matter (see, e.g.,
\cite{pion,migdal90}).
If we first neglect the effect of strong correlations of pions
with the matter in modifying the pion self-energy, one finds
it is favorable for a neutron on the top of the Fermi sea to 
turn into a proton and a $\pi^-$ when 
\begin{equation}
    \mu_n-\mu_p=\mu_e > m_{\pi^-},
\end{equation}
where $m_{\pi^-}=139.6$ MeV is the $\pi^{-}$ rest mass. As discussed 
in the previous subsection, at nuclear matter 
saturation density the electron chemical potential is
$\sim 100$ MeV and one might therefore expect the appearance
of $\pi^{-}$ at a slightly higher density.
One can however not neglect the interaction of the pion with the 
background matter. Such interactions can enhance the pion self-energy
and thereby the pion threshold density, and depending on the chosen
parameters, see again Ref.\ \cite{migdal90}, the critical density 
for pion condensation may vary from $n_0$ to $4n_0$.  
These matters are however
not yet settled in a satisfying way, and models with strong nucleon-nucleon
correlations tend to suppress both the $\pi NN$ and $\pi\Delta N$ 
interaction vertices so that a pion condensation in neutron star matter
does not occur. 

A $\pi^0$ condensate may also form as recently suggested by Akmal et
al.\ \cite{Akmal} and appears at a density of $\sim 0.2$ fm$^{-3}$ for
pure neutron matter when the three-body interaction is included,
whereas without $V_{ijk}$ it appears at much higher densities, i.e.\
$\sim 0.5$ fm$^{-3}$. The $\pi^0$ are virtual in the same way as the
photons keeping a solid together are virtual.

\subsection{Hyperon matter}

Condensates of  hyperons $\Lambda,\Sigma^{-,0,+},...$ and Deltas
$\Delta^{-,0,+,++}$ also appear when
their chemical potential exceeds their effective mass in matter.  
In $\beta$-equilibrium the chemical potentials are related by
\bea
   \mu_n+\mu_e &=& \mu_{\Sigma^-} =\mu_{\Delta^-} \\
   \mu_n &=& \mu_{\Lambda} =  \mu_{\Sigma^0} = \mu_{\Delta^0} \\
   \mu_p &=& \mu_{\Delta^+} \,.
\eea
The $\Sigma^-$ appears via weak strangeness non-conserving
interactions $e^-+n \rightarrow \Sigma^- +\nu_e$, when $\mu_{\Sigma^-}
> \omega_{\Sigma^-}$ and $\Lambda$ hyperons when $\mu_{\Lambda}>
\omega_{\Lambda}$ If we neglect interactions, one would expect the
$\Sigma^-$ to appear at lower densities than the $\Lambda$, even
though $\Sigma^-$ is more massive due to the large electron chemical
potential.  The threshold
densities for noninteracting $\Sigma^-,\Delta^-,\Lambda$ are
relatively low indicating that these condensates are present in cores
of neutron stars if their interactions can be ignored.

Hyperon energies are, however, strongly affected in dense nuclear
matter by interactions and correlations with the nucleons and
other hyperons in a condensate. If hyperons have the short range
repulsion and three-body interactions as nucleons, a condensate
of hyperons become less likely.

\subsection{Quark matter}

Eventually at high densities, we expect hadrons to
deconfine to
quark matter or, in other words, chiral symmetry to be restored.
This transition has been extensively studied by the Bag model
equation of state (EoS) for quark matter which leads to
a first order phase transition from hadronic to quark matter at 
a density and temperature determined by the parameters going into
both EoS.
In the bag model the quarks are assumed to be confined to
a finite region of space, the so-called 'bag', by a vacuum pressure
$B$.  Adding the
Fermi pressure and interactions computed to order $\alpha_s=g^2/4\pi$,
where $g$ is the QCD coupling constant, the total pressure
for three massless quarks of flavor $f=u,d,s$, is 
\begin{equation}
    P=\frac{3\mu_f^4}{4\pi^2}(1-\frac{2}{\pi}\alpha_s) -B +P_e+P_\mu \,,
     \label{pquark}
\end{equation}
where $P_{e,\mu}$ are the electron and muon pressure, e.g., 
$P_e=\mu_e^4/12\pi^2$.
A Fermi gas of quarks of flavor {\em i} has density $n_i =
k_{Fi}^3/\pi^2$, due to the three color states. 
A finite strange quark mass have minor effect on the EoS since
quark chemical potentials $\mu_q\ga m_N/3$ typically are
much larger. The value of the bag constant {\em B} is poorly
known, and we present results using two representative values,
$B=150$ MeVfm$^{-3}$ and $B=200$ MeVfm$^{-3}$.
We take $\alpha_s=0.4$. However, similar results can be obtained with
smaller $\alpha_s$ and larger $B$.

\begin{figure}
\begin{center}
{\centering\mbox{\psfig{file=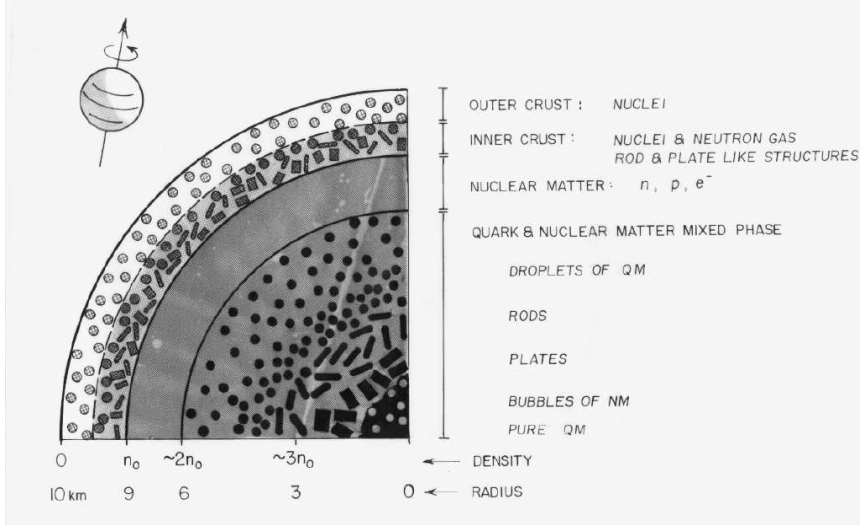,height=7cm,angle=90}}}
\vspace{0cm}
\caption{The quark and nuclear matter structure in a quarter of a typical
1.4$M_\odot$ solar mass neutron star.
The typical sizes of structures are a few Fermi's but have been scaled
up by about 16 orders of magnitudes to be seen.  \label{star}  }
\end{center}
\end{figure}

 The quark and nuclear matter mixed phase has continuous pressures and
densities due to the general Gibbs criteria for two-component systems
\cite{Glendenning}. There are no first order phase transitions but at
most two second order phase transitions. Namely, at a lower density,
where quark matter first appears in nuclear matter, and at a very high
density (if gravitationally stable), where all nucleons are finally
dissolved into quark matter. This mixed phase does, however, not
include local surface and Coulomb energies of the quark and nuclear
matter structures. If the interface tension between quark and nuclear
matter is too large, the mixed phase is not favored energetically due
to surface and Coulomb energies associated with forming these
structures \cite{HPS}. The neutron star
will then have a core of pure quark matter with a mantle of nuclear
matter surrounding it and the two phases are coexisting by a first
order phase transition or Maxwell construction.  For a small or
moderate interface tension the quarks are confined in droplet, rod-
and plate-like structures as found in the inner crust of neutron stars
(see \cite{lrp93} and Fig. \ref{star}).

\subsection{Superfluidity in baryonic matter}

The presence of neutron superfluidity in the crust and the inner part
of neutron stars are considered well established in the physics of
these compact stellar objects.  In the low density outer part of a
neutron star, the neutron superfluidity is expected mainly in the
attractive $^1S_0$ channel.  At higher density, the nuclei in the
crust dissolve, and one expects a region consisting of a quantum
liquid of neutrons and protons in beta equilibrium.  The proton
contaminant should be superfluid in the $^1S_0$ channel, while neutron
superfluidity is expected to occur mainly in the coupled
$^3P_2$-$^3F_2$ two-neutron channel.  In the core of the star any
superfluid phase should finally disappear.

Dilute Fermi systems can now be studied as atomic gases
recently have been cooled down to nanokelvin similar to Bose-Einstein
condensates. Degeneracy was observed
\cite{Jin} and BCS gaps are currently
searched for. According to Gorkov and Melik-Bharkuderov \cite{Gorkov} the
gap is
\bea
  \Delta=\left(\frac{2}{e}\right)^{7/3} \frac{k_F^2}{2m} 
   \exp\left[\frac{\pi}{2ak_F}\right] \, \label{gap} 
\eea
in the dilute limit where the Fermi momentum times the scattering length
is small, $k_F|a|\ll1$.
 
Recently color superconductivity in quark matter has been taken up
again since the quark color interaction has been conjectured to be
large.  Correspondingly large gaps of order tens of MeV are found.  If
the strange quark mass is sufficiently small color-flavor locking
occur between up and down or strange quarks.  We refer to T. Schaefer
these proceedings for further details.

\section{Calculated neutron star masses and radii}

In order to obtain the mass and radius of a neutron star, we have solved the
Tolman-Oppenheimer-Volkov equation with and without rotational
corrections.  The equations of
state employed are given by the $pn$-matter EoS with $s =0.13,
0.2, 0.3, 0.4$ with nucleonic degrees of freedom only. In addition we
have selected two representative values for the bag-model parameter
$B$, namely 150 and 200 MeVfm$^{-3}$ for our discussion on eventual phase
transitions. The quark phase is linked with our $pn$-matter EoS from
Eq.\ (\ref{eq:EA}) with $s=0.2$ through either a mixed phase
construction or a Maxwell construction \cite{physrep}.  
For $B=150$ MeVfm$^{-3}$, the
mixed phase begins at 0.51 fm$^{-3}$ and the pure quark matter phase
begins at $1.89$ fm$^{-3}$.  Finally, for $B=200$ MeVfm$^{-3}$, the mixed
phase starts at $0.72$ fm$^{-3}$ while the pure quark phase starts
at $2.11$ fm$^{-3}$.  In case of a Maxwell construction, in order to
link the $pn$ and the quark matter EoS, we obtain for $B=150$ MeVfm$^{-3}$
that the pure $pn$ phase ends at $0.92$ fm$^{-3}$ and that the
pure quark phase starts at $1.215$ fm$^{-3}$, while the corresponding
numbers for $B=200$ MeVfm$^{-3}$ are $1.04$ and $1.57$
fm$^{-3}$.

None of the equations of state
from either the pure $pn$ phase or with a mixed phase or Maxwell
construction with quark degrees of freedom, result in stable
configurations for densities above $\sim 10 n_0$, implying thereby
that none of the stars have cores with a pure quark phase.  The EoS
with $pn$ degrees of freedom have masses $M\la2.2M_{\odot}$ when
rotational corrections are accounted for.  With the inclusion of the
mixed phase, the total mass is reduced since the EoS is softer.
 However, there is the possibility of
making very heavy quark stars for very small bag constants. For pure
quark stars there is only one energy scale namely $B$ which provides
a homology transformation \cite{Madsen} and the maximum mass is
$M_{max}=2.0M_\odot (58{\rm MeV fm^{-3}}/B)^{1/2}$ (for
$\alpha_s=0$). However, for $B\ga 58{\rm MeV fm^{-3}}$ a nuclear
matter mantle has to be added and for $B\la 58{\rm MeV fm^{-3}}$ quark
matter has lower energy per baryon than $^{56}$Fe and is thus the
ground state of strongly interacting matter. Unless the latter is the
case, we can thus exclude the existence of such $2.2-2.3M_\odot$ quark
stars.

\begin{figure}
\vspace{-1cm}
\begin{center}
{\centering\mbox{\psfig{file=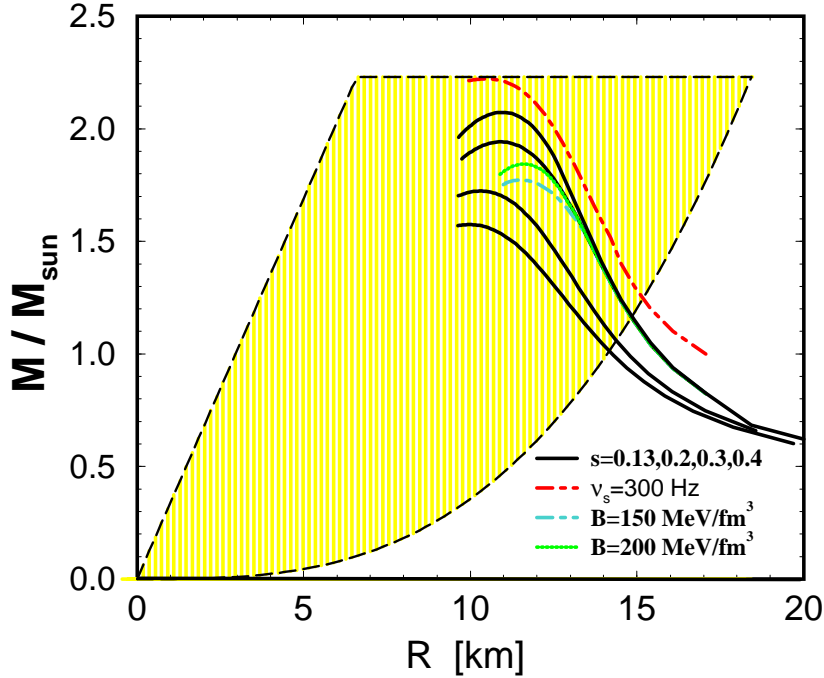,height=10cm,angle=0}}}
\caption{Neutron star masses vs.~radius for the EoS of Eq.~(\ref{eq:EA}) 
with softness s=0.13,0.2,0.3,0.4, with increasing values of $s$ 
from top to bottom for the full curves.
Phase transitions decrease the maximum mass whereas rotation
increases it. The shaded area represents  the neutron star
radii and masses allowed (see text and Eqs. 1-3) for 
orbital QPO frequencies 1060~Hz of 4U 1820-30. 
\label{fig4}  }
\end{center}
\vspace{-0.5cm}
\end{figure}

In Fig. \ref{fig4} we show the mass-radius relations for the various
equations of state. 
The shaded area represents the allowed masses and radii for 
$\nu_{QPO}=1060$ Hz of 4U 1820-30. Generally,
\begin{eqnarray}
  2GM < R < \left(\frac{GM}{4\pi^2\nu_{QPO}^2}\right)^{1/3} \,,
\end{eqnarray}
where the lower limit insures that the star is not a black hole,
and the upper limit that the accreting matter orbits outside
the star, $R<R_{orb}$. Furthermore,
for the matter to be outside the innermost
stable orbit, $R>R_{ms}=6GM$, requires that 
\begin{eqnarray}
   M &\la& \frac{1+0.75j}{12\sqrt{6}\pi G\nu_{QPO}} \,
 \simeq 2.2 M_\odot (1+0.75j)\frac{{\rm kHz}}{\nu_{QPO}} \,,  \label{Mms} 
\end{eqnarray}
where $j=2\pi c\nu_sI/M^2$ is a dimensionless measure of the angular
momentum of the star with moment of inertia $I$.  The upper limit
in Eq. (\ref{Mms}) is the mass when $\nu_{QPO}$ corresponds to the innermost
stable orbit. This is the case for 4U 1820-30 since according to \cite{Zhang}
$\nu_{QPO}$ saturates at $\sim1060$~Hz with increasing count rate.
The corresponding neutron star mass is $M\sim 2.2-2.3M_\odot$ which
leads to several interesting conclusions as seen in Fig.\
\ref{fig4}. Firstly, the stiffest EoS allowed by causality
($s\simeq 0.13-0.2$) is needed. Secondly, rotation must be included
which increase the maximum mass and corresponding
radii by 10-15\% for $\nu_s\sim 300$~Hz.
Thirdly, a phase transition to quark matter below densities of order
$\sim 5 n_0$ can be excluded, corresponding to a restriction on the bag
constant $B\ga200$ MeVfm$^{-3}$.

 These maximum masses are smaller than those of APR98 and Kalogera \&
Baym who, as discussed above, obtain upper bounds on the mass of
neutron stars by discontinuously setting the sound speed to equal the
speed of light above a certain density, $n_c$. By varying the density
$n_c=2\to 5n_0$ the maximum mass drops from $2.9\to 2.2M_\odot$. In
our case, incorporating causality smoothly by introducing the
parameter $s$ in Eq.\ (\ref{eq:EA}), the EoS is softened at higher
densities in order to obey causality, and yields a maximum mass 
lower than the $2.2M_\odot$ derived in APR98 for nonrotating
stars.

If the QPOs are not from the innermost stable orbits and one finds
that even accreting neutron stars have small masses, say like the
binary pulsars, $M\la1.4M_\odot$, this may indicate that heavier
neutron stars are not stable. Therefore, the EoS is soft at high
densities $s\ga0.4$ or a phase transition occurs at a few
times nuclear matter densities.

\section{Glitches and phase transitions in rotating neutron stars}

 Younger pulsars rotate and slow down rapidly. Some display
sudden speed ups referred to as glitches.
The glitches observed in the Crab, Vela, and a few other pulsars are
probably due to quakes occurring in solid structures such as the
crust, superfluid vortices or possibly the quark matter lattice in the
core \cite{HPS}.  As the rotating neutron star gradually slows down and
becomes less deformed, the rigid component is strained and eventually
cracks/quakes and changes its structure towards being more spherical.

The moment of inertia of the rigid component, $I_c$, decreases
abruptly and its rotation and pulsar frequency increases due to
angular momentum conservation resulting in a glitch.  The observed
glitches are very small $\Delta\Omega/\Omega\sim 10^{-8}$.  The two
components slowly relaxate to a common rotational frequency on a
time scale of days (healing time) due to superfluidity of the other
component (the neutron liquid).  The {\it healing parameter}
$Q=I_c/I_{tot}$ measured in glitches reveals that for the Vela and
Crab pulsar about $\sim$3\% and $\sim$96\% of the moment of inertia is
in the rigid component respectively.

If the crust were the only rigid component the Vela neutron star
should be almost all crust.  This would require that the Vela is a
very light neutron star - much smaller than the observed ones which
all are compatible with $\sim 1.4M_\odot$.  If we by the lattice
component include not only the solid crust but also the protons in 
nuclear matter (NM)
(which is locked to the crust due to magnetic fields), superfluid
vortices pinned to the crust and the solid QM mixed phase
\begin{equation}
   I_c = I_{crust}+I_p+I_{sv}+I_{QM} \, ,
\end{equation}
we can better explain the large $I_c$ for the Crab.  The moment of
inertia of the mixed phase is sensitive to the EoS's used.  For
example, for a quadratic NM EoS \cite{HPS} decreasing the Bag constant
from 110 to 95 MeVfm$^{-3}$ increases $I_c/I_{total}$ from $\sim20\%$
to $\sim70\%$ for a 1.4$M_\odot$ neutron star - not including possible
vortex pinning.  The structures in the mixed phase would exhibit
anisotropic elastic properties, being rigid to some shear strains but
not others in much the same way as liquid crystals. Therefore the
whole mixed phase might not be rigid.

As the neutron star slows down, pressures and densities increase in
the center and a first order phase transition may occur. 
Does it leave any detectable signal at the corresponding {\it critical}
angular velocity $\Omega_0$ ?
As described in detail in \cite{HJ,physrep} the general relativistic 
equations for slowly rotating stars can be solved even with
first order phase transitions since only the monopole is important. 
The resulting moment of inertia
have the characteristic behavior for 
$\Omega \raisebox{-.5ex}{$\stackrel{<}{\sim}$}\Omega_0$
\begin{equation}
  I = I_0\left( 1 +c_1\Omega^2-c_2(\Omega_0^2-\Omega^2)^{3/2} + ... \right) . 
  \label{Igen}
\end{equation}
Here, $c_1$ and $c_2$ are small parameters 
proportional to the density difference
between the two phases; however, $c_2=0$ for $\Omega>\Omega_0$. 

In order to make contact with observation, the temporal behavior
of angular velocities must be considered. The pulsars slow down at a rate
given by the loss of rotational energy which one usually assumes is
proportional to the rotational angular velocity to some power
(for dipole radiation $n=3$)
\begin{equation}
  \frac{d}{dt} \left(\frac{1}{2}I\Omega^2\right) = -C \Omega^{n+1}. 
   \label{dE}
\end{equation}
With the moment of inertia given by Eq. (\ref{Igen})
the decreasing angular velocity can be found.
The corresponding 
braking index depends on the second derivative $I''=dI/d^2\Omega$
of the moment of inertia and thus diverges
as $\Omega$ approaches $\Omega_0$ from below
\begin{eqnarray}
     n(\Omega) &\equiv& \frac{\ddot{\Omega}\Omega}{\dot{\Omega}^2} 
    \simeq n - c_1\Omega^2
    +c_2\frac{\Omega^4}{\sqrt{\Omega_0^2-\Omega^2}} \,.\label{n}
\end{eqnarray}
The  {\it observational} braking index $n(\Omega)$ 
exhibits a characteristic behavior
different from the {\it theoretical} braking index $n$ in case of a 
first order phase transition.

The critical angular velocities depend on the EoS and the critical
densities. For best detection one would want the transition to occur
for rapidly rotating pulsars such as millisecond pulsars, X-ray
binaries or young neutron stars only a few years or centuries old.  As
pulsars slow down over a million years, their central densities span a
wide range of order several $n_0$. As we are interested in time scales
of years, we must instead study the $\sim 1000$ pulsars available.  By
studying the corresponding range of angular velocities for the sample
of different star masses, the chance for encountering a critical
angular velocity increases.  Eventually, one may be able to cover the
full range of central densities and find all first order phase
transitions up to a certain size determined by the experimental
resolution.

\section{Other recent surprises related to neutron stars}

\subsection{Connecting $NO^-_3$ peaks in ice core samples to supernovae}

A curious connection between four $NO^-_3$ peaks from south pole ice
core samples and supernovae has recently been made.  A small nearby
but well hidden supernova remnant RX J0852.0-4622 was discovered
\cite{SN4} in the southeast corner of the older Vela supernova
remnant.  Estimates of its distance and age seem to be compatible with
the 4th $NO^-_3$ peak in the 20 year old south pole core samples
\cite{ice}.  The other three already agreed with the historical Crab,
Tycho and Kepler supernovae within the ice dating accuracy of $\pm20$
years, a millennium back.  This seems to indicate that nearby supernova
explosions can affect the climate on earth and lead to geophysical
signals. The radiation does produce $NO^-_3$ in the upper atmosphere
but the quantitative amount of fall out on the poles by
northern/southern light cannot be estimated.  $NO^-_3$ peaks were not
found in the Vostok og Greenland drills possibly because the aurorae
do not have fallout on these sites. Further drillings are needed to
confirm the connection - in particular deeper ones which should also
include the 1006 and 1054 supernovae. Mass extinctions may be caused
by nearby supernova explosions after all.

\subsection{X-ray bursts and thermonuclear explosions}

Slow accretion from a small mass companion $\la 2M_\odot$ generates a
continuous background of X-rays. Each nucleon radiates its
gravitational energy of $\sim m_nGM/R\simeq 100$~MeV. After
accumulating hydrogen on the surface, pressures and temperatures
become sufficient to trigger an irregular runaway thermonuclear
explosion every few hours or so seen as an X-ray burst (see, e.g.,
\cite{Paradijs} for a review). The energy involved is typical nuclear
binding energies $\sim 1$~MeV, i.e., a percent of the time integrated
background.

 The microsecond time resolution in these X-ray spectra allow for
Fourier transforms in millisecond time intervals which is much shorter
than the burst duration of order a few seconds.  In the case of 4U
1728-34 the power analysis shows a peak at the neutron star spin
frequency at 364~Hz which, however, decreases to 362~Hz during the
first 1-2~seconds of the burst \cite{PhysicsToday}. A simple
explanation is that the thermonuclear explosion elevates the surface
of the neutron star.  Conserving angular momentum, $L\propto MR^2\nu$,
leads to a decrease in rotation by
\be 
   \frac{\Delta\nu}{\nu} \simeq -2\frac{\Delta R}{R} \,.  
\ee
With a frequency change of  $\Delta\nu\sim -2$~Hz and typical neutron
star radii of order $R\sim 10$~km, we find an elevation of order
$\Delta R\sim 20$~m, which is roughly in agreement
with expectations but much less than on earth due
to the much stronger gravitational fields on neutron stars.

\subsection{Gamma Ray Bursters and hypernovae}

The recent discovery of afterglow in Gamma Ray Bursters (GRB) allows
determination of the very high redshifts ($z\ge 1$) and thus the
enormous distance and energy output $E\sim 10^{53}$ ergs in GRB if
isotropically emitted. Very recently evidence for beaming or jets has
been found \cite{Kulkarni} corresponding to ``only'' $E\sim 10^{51}$
ergs. Candidates for such violent events include neutron star mergers or a
special class of type Ic supernova ({\it hypernovae}) where cores
collapse to black holes. Indication of such connections are brought by
recent observations of a bright supernova coinciding with GRB 980326.
Binary pulsars are rapidly spiraling inwards and will
eventually merge and create a gigantic explosion and perhaps collapse
to a black hole. From the number of binary pulsars and spiral rates,
one estimates about one merger per million year per galaxy. With $\sim
10^9$ galaxies at cosmological distances $z\la 1$ this gives about the
observed rate of GRB.  However, detailed calculations have problems
with baryon contamination, i.e.\  the baryons ejected absorb photons in
the relativistically expanding photosphere. Accreting black holes have
also been suggested to act as beaming GRB.

So far, the physics producing these GRB is not understood.  The time
scales and the enormous power output points towards neutron star or
black hole objects. 

\section{Summary}

Modern nucleon-nucleon potentials have reduced the uncertainties in
the calculated EoS.  Using the most recent realistic effective
interactions for nuclear matter of APR98 with a smooth extrapolation
to high densities including causality, the EoS could be
constrained by a ``softness'' parameter $s$ which parametrizes the
unknown stiffness of the EoS at high densities. Maximum masses have
subsequently been calculated for rotating neutron stars with and
without first and second order phase transitions to, e.g., quark
matter at high densities.
The calculated bounds for maximum masses leaves two natural options
when compared to the observed neutron star masses:

\begin{itemize}
 \item {\bf Case I}: {\it The large masses of the neutron stars in
QPO 4U 1820-30 ($M=2.3M_\odot$), PSR J1012+5307
($M=2.1\pm0.4 M_\odot$), Vela X-1 ($M=1.9\pm0.1 M_\odot$), and
Cygnus X-2 ($M=1.8\pm0.2 M_\odot$), are confirmed and
complemented by other neutron stars with masses around $\sim 2M_\odot$.}
The EoS of dense nuclear matter is then severely
constrained and only the stiffest EoS consistent with causality are allowed,
i.e., softness parameter $0.13\le s\la0.2$.  Furthermore, any
significant phase transition at densities below $< 5n_0$ can be
excluded. 

 That the radio binary pulsars all have masses around $1.4M_\odot$ is
then probably due to the formation mechanism in supernovae where the
Chandrasekhar mass for iron cores are $\sim1.5M_\odot$.
Neutron stars in binaries can subsequently acquire larger
masses by accretion as X-ray binaries.

 \item {\bf Case II}: 
{\it The heavy neutron stars prove erroneous by more detailed observations
and only masses like those of binary pulsars are found.}
If accretion does not produce neutron stars heavier than 
$\ga1.4M_\odot$, this indicates that heavier neutron stars simply are not
stable which in turn implies a soft EoS, either $s> 0.4$ or a
significant phase transition must occur already at a few times nuclear
saturation densities. 

\end{itemize}

Surface temperatures can be estimated from spectra and from the
measured fluxes and known distances, one can extract the surface
area of the emitting spot. This gives unfortunately only a lower limit
on the neutron star size, $R$. If it becomes possible to measure
both mass and radii of neutron stars, one can plot an observational
$(M,R)$ curve in Fig. (\ref{fig4}), which uniquely determines the
EoS for strongly interacting matter at low temperature.

Pulsar rotation frequencies and glitches are other promising signals
that could reveal phase transitions. Besides the standard glitches
also giant glitches were mentioned and in particular the
characteristic behavior of angular velocities when a first order
phase transition occurs right in the center of the star.

It is impossible to cover all the interesting
and recent developments concerning neutron stars in these proceedings
so for more details we refer to \cite{physrep} and Refs. therein.

\vspace*{-2pt}

\section*{Acknowledgments}
Thanks to my collaborators M. Hjorth-Jensen
and C.J.~Pethick as well as G. Baym, F. Lamb and V. Pandharipande.

\eject

\end{document}